\documentclass{article}
\usepackage{spconf,amsmath,graphicx}
\usepackage{amssymb}
\usepackage{amsmath}
\usepackage{url}
\usepackage[OT1]{fontenc}
\usepackage{xcolor}

\title{COMBOLOSS FOR FACIAL ATTRACTIVENESS ANALYSIS WITH SQUEEZE-AND-EXCITATION NETWORKS}
\name{Lu Xu and Jinhai Xiang\sthanks{Corresponding author.}}
\address{College of Informatics, Huazhong Agricultural University, Wuhan, 430070, China}

\begin{document}
%\ninept
%
\maketitle
\begin{abstract}
Loss function is crucial for model training and feature representation learning, conventional models usually regard facial attractiveness recognition task as a regression problem, and adopt MSE loss or Huber variant loss as supervision to train a deep convolutional neural network (CNN) to predict facial attractiveness score. Little work has been done to systematically compare the performance of diverse loss functions. In this paper, we firstly systematically analyze model performance under diverse loss functions. Then a novel loss function named \textit{ComboLoss} is proposed to guide the SEResNeXt50 network. The proposed method achieves state-of-the-art performance on SCUT-FBP, HotOrNot and SCUT-FBP5500 datasets with an improvement of 1.13\%, 2.1\% and 0.57\% compared with prior arts, respectively. Code and models are available at ~\url{https://github.com/lucasxlu/ComboLoss.git}.
\end{abstract}
\begin{keywords}
Deep learning, facial beauty prediction (FBP), face analysis
\end{keywords}
\section{Introduction and Related Work}
\label{sec:intro}
With the popularity of short video platforms (such as TikTok~\footnote{\url{https://www.tiktok.com}}) and social network APPs (like Facebook, Instagram and WeChat) among the public, facial attractiveness analysis gains increased attention. Previous works~\cite{perrett1994facial,Eisenthal2006Facial,gray2010predicting} demonstrate that computational models can be adopted to automatically learn facial attractiveness. Recent years have witnessed many achievements in related areas~\cite{xie2015scut,xu2018transferring,xu2018crnet,xu2019hierarchical,lin2019attribute}. However, due to the diverse head poses, expression, age, low resolution and illumination problems, it is still quite challenging to develop an accurate model to predict facial attractiveness levels.

Facial attractiveness analysis has been researched for decades with many achievements~\cite{perrett1994facial,Eisenthal2006Facial,kagian2007humanlike,liang2018scut,gray2010predicting,xu2018crnet,Xu2017Facial,xu2019hierarchical}. Traditional methods~\cite{perrett1994facial,xie2015scut} usually extract hand-crafted features to form facial representation, and train a regressor or a classifier. The output of the model is regarded as the facial attractiveness level. Gray et al.~\cite{gray2010predicting} introduce a shallow neural network to learn facial beauty on their proposed HotOrNot dataset~\cite{gray2010predicting} without utilizing facial landmarks. Since the breakthrough of AlexNet~\cite{krizhevsky2012imagenet}, researchers are paying more attention to developing more advanced CNN architecture to improve facial beauty prediction (FBP) accuracy. Liang et al.~\cite{liang2018scut} introduce SCUT-FBP5500 dataset~\cite{liang2018scut} with 5500 portrait images, which facilitate further research~\cite{xu2019hierarchical,lin2019attribute,lin2019regression} in related fields. Xu et al.~\cite{xu2018crnet} optimize a classification branch and a regression branch in parallel with their proposed CRNet~\cite{xu2018crnet}, and report very promising results on SCUT-FBP~\cite{xie2015scut} and HotOrNot~\cite{gray2010predicting} dataset, but the exploration of the classification branch is limited. Lin et al.~\cite{lin2019attribute} propose AaNet which takes beauty-related facial attributes as additional inputs to enhance model performance. Despite the promising performance, the subnetworks of auxiliary tasks induce additional network parameters, which makes the facial attractiveness analysis model quite heavy. In contrast to existing works~\cite{xu2019hierarchical,xu2018crnet,lin2019attribute}, we do not add any heavy extra subnetworks into the backbone model, result in less probability of overfitting. By simply better utilizing the network's output, we append a classification loss (aka weighted cross entropy loss), an expectation loss and a regression loss to better supervise the model training. We term our approach as \textit{ComboLoss}, experimental results indicate that ComboLoss achieves state-of-the-art performance on 3 datasets without bells and whistles.

The main contributions of this paper are as follows: (1) We systematically compare different loss functions in guiding a deep CNN to learn facial attractiveness. (2) We present a simple yet effective approach named ComboLoss, to better leverage the output of CNN and enhance the performance as well, while result in neglectable (less than 0.037\%) parameters increasement. (3) The proposed methods achieve state-of-the-art performance on SCUT-FBP~\cite{xie2015scut}, HotOrNot~\cite{gray2010predicting} and SCUT-FBP5500~\cite{liang2018scut} datasets, surpassing prior arts by 1.13\%, 2.1\% and 0.57\%, respectively.

\section{Proposed Methods}
\label{sec:methods}

\subsection{ComboLoss}
In this section, we give a detailed discussion of our proposed ComboLoss. The target of deep CNN-based facial attractiveness analysis is to find a nonlinear mapping $y=\Phi(\mathcal{I})$, which means a deep CNN $\Phi$ maps an input facial image $\mathcal{I}\in \mathcal{R}^{W\times H\times C}$ to an output beauty score $y$. Given a set of labelled groundtruth training samples $\Omega=\{\mathcal{I}_i, \hat{y}_i\}_{i=1}^N$, the target of the model training is to find an optimal $\Phi$ which minimizes:
\begin{equation}
	\sum_{i=1}^{N}\mathcal{L}(\Phi(\mathcal{I}_i), \hat{y}_i)
\end{equation}
where $\mathcal{L}$ is a pre-defined loss function which measures the difference between a predicted beauty score and its groundtruth beauty score. In this phenomenon, the deep CNN is regarded as a regression model.

The design of a proper loss function is crucial for deep CNN-based facial beauty prediction. However, the majority deep CNN-based FBP systems~\cite{xie2015scut,liang2018scut,Xu2017Facial} usually utilize MSE as supervision. The performance of solely leveraging beauty score as supervision to train a regression network is limited~\cite{xu2018crnet,xu2019hierarchical,lin2019regression}, researchers are paying attention to adopt multi-task learning~\cite{xu2019hierarchical}, ranking~\cite{lin2019regression}, multi-stream network~\cite{xu2018crnet}, label distribution learning~\cite{liu2017facial}, and utilize auxiliary network~\cite{lin2019attribute} to enhance performance. However, these methods bring too many parameters, result in quite heavy models. In this paper, we propose a new loss function, termed \textit{ComboLoss}, which further improves the performance of deep CNN-based FBP systems. The components of ComboLoss are regression loss, expectation loss, and a classification loss.

\begin{itemize}
	\item \textbf{Regression Loss}.
	In contrast to mainstream FBP models with MSE loss, we adopt $L_1$ loss as regression loss instead. The regression loss is defined as:
	\begin{equation}
		L_{reg}=\frac{1}{N}\sum_{i=1}^{N}|\hat{s}_i-s_i|
	\end{equation}
	where $\hat{s}_i, s_i$ and $N$ represent the output score of regression module, groundtruth score and sample capacity, respectively.
	\item \textbf{Classification Loss}.
	In addition to regression module, we also insert a classification module before the last layer of SEResNeXt50 network. To enhance loss and gradient flowing, we adopt weighted cross entropy as classification loss, which is defined as~Equation~\ref{eq:cls}. The chosen of output neuron $\mathcal{C}$ and $w_c$ are discussed in Section~\ref{subsec:discretization}. We introduce classification loss into a regression network to learn additional information, to better separates attractive faces and unattractive samples.
	\begin{equation}
	\label{eq:cls}
		L_{cls}=-\frac{1}{N}\sum_{i=1}^{N}w_c c_ilog\hat{c}_i
	\end{equation}
	where $\hat{c}_i$, $c_i$ and $w_c$ denote predicted probability output by softmax layer, correct classification indicator and weight of each category, respectively.
	\item \textbf{Expectation Loss}.
	Despite the classification loss mentioned above can better separates different facial attractive levels, but it holds the assumption that categories are independent of each other, which is not suitable in this phenomenon (e.g. a face with a beauty level of 4 is undoubtedly more attractive than a face with beauty level of 1). In order to enhance model training, we present a new loss function by leveraging softmax probability. Then a regression manner is applied to minimize the gap between groundtruth beauty score and the expectation score. We define \textit{expectation} as:
	\begin{equation}
		\mathcal{E}_i=\sum_{i=1}^{\mathcal{C}}\hat{c}_i\cdot i
	\end{equation}
	The \textit{expectation loss} measures the difference between groundtruth score and expectation score with $L_1$ loss:
	\begin{equation}
		L_{exp}=\frac{1}{N}\sum_{i=1}^{N}|\hat{s}_i-\mathcal{E}_i|
	\end{equation}
\end{itemize}

Having defined the individual item, the ComboLoss is denoted as $\mathcal{L}_{combo}=\alpha \cdot L_{reg} + \beta \cdot L_{exp} + \gamma\cdot L_{cls}$, we set $\alpha=2,\beta=1,\gamma=1$ in our experiments.

\subsection{Attractiveness Score Discretization}
\label{subsec:discretization}
Since we insert \textit{classification module} into a regression network, the groundtruth for classification should be defined for model training. The groundtruth beauty scores in relevant benchmark dataset~\cite{xie2015scut,liang2018scut,gray2010predicting} are continuous values, which cannot be directly utilized to train a classification task. Therefore, attractiveness score discretization should be conducted firstly. How to discrete continuous scores and how many score ranges should be discreted remain a open problem. In this paper, we take the simplest fashion for easy implementation. Formally, in SCUT-FBP~\cite{xie2015scut} and SCUT-FBP5500~\cite{liang2018scut} datasets, the rounded integer of $\lceil s-\frac{1}{2}\rceil$ is regarded as classification label, in HotOrNot~\cite{gray2010predicting}, the intervals are set to 3 as in~\cite{xu2018crnet}. More advanced discretization methods may also be adopted, which is left to our future work.

It is worth noting that despite we take the similar discretization method like CRNet~\cite{xu2018crnet}, but the imbalance problem after applying discretization is neglected in \cite{xu2018crnet}. The capacity of category 0 is 3 times bigger than category 1, which is harmful for model training. In this paper, we solve the imbalanced learning problem in loss level. Namely, the hyper-parameter in Equation~\ref{eq:cls} is introduced to balance the loss and gradient of each category. The hyper-parameter $w_c$ of category $c\in \mathcal{C}$ is determined as:
\begin{equation}
	w_c=\frac{max(|m|)_{m\in \mathcal{C}}}{|c|}
\end{equation}

\subsection{Network Architecture}
Inspired by \cite{hu2018squeeze}, we incorporate squeeze-and-excitation module~\cite{hu2018squeeze} into the ResNeXt50~\cite{xie2017aggregated} to form SEResNeXt50 as the backbone network for feature representation learning. Other architecture (such as ResNet~\cite{he2016deep}, EfficientNet~\cite{tan2019efficientnet}) may also be utilized, but it's beyond the scope of this paper.

\begin{figure}[htbp]
	\includegraphics[width=0.4\textheight]{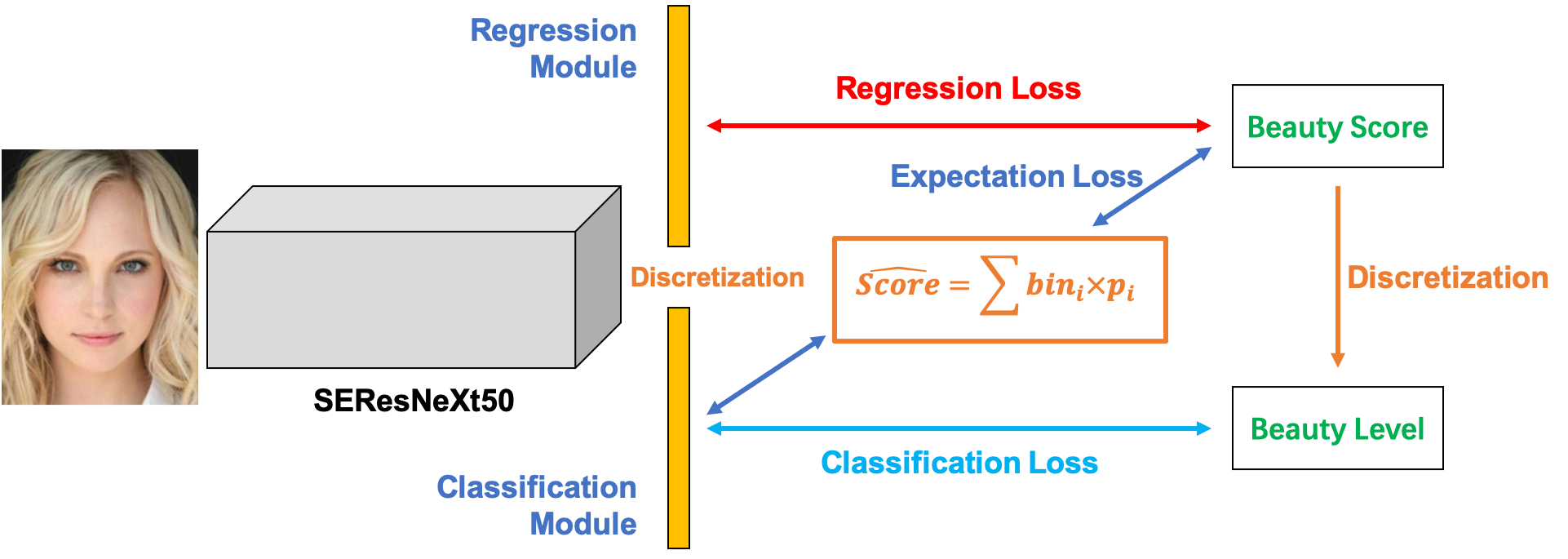}
	\centering
	\caption{Network architecture supervised by ComboLoss.}
	\label{fig:comboloss}
\end{figure}

Squeeze operation can embed global spatial information with a channel descriptor through utilizing global average pooling to generate channel-wise statistics. Namely, by global average pooling the feature map, we get:
\begin{equation}
	z_c=\frac{1}{H\times W}\sum_{i=1}^{H}\sum_{j=1}^{W} u_c(i,j)
\end{equation}
where $H$ and $W$ represent the height and width of the feature map, $u_c(i,j)$ denotes the pixel value at $(i,j)$ in channel $c$. After obtaining the statistics $z_c\in \mathbb{R}^C$, an excitation operation is applied. This is achieved by the following equation:
\begin{equation}
	s=\sigma (W_2\cdot ReLU(W_1\cdot z))
\end{equation}
where $\sigma$ denotes sigmoid activation, $W_1\in \mathbb{R}^{\frac{C}{r}\times C}$ is a dimensionality-reduction fully connected (FC) layer with reduction ratio $r$, and $W_2\in \mathbb{R}^{C\times \frac{C}{r}}$ is a dimensionality-increasing FC layer with increasing ratio $r$. The output of the squeeze-and-excitation block is calculated by rescaling the transformation output $U$ with activations:
\begin{equation}
	\tilde{x}_c=s_c\cdot u_c
\end{equation}
where $\tilde{X}=[\tilde{x}_1,\cdots,\tilde{x}_c]$, $u_c\in \mathbb{R}^{H\times W}$ represents feature map, $s_c$ represents scalar of sigmoid activation output (please refer to~\cite{hu2018squeeze} for more details). In this paper, we incorporate squeeze-and-excitation~\cite{hu2018squeeze} module into ResNeXt50~\cite{xie2017aggregated}, result in SEResNeXt50 architecture. The network architecture is shown in Fig~\ref{fig:comboloss}.

\section{Experiments}
\label{sec:experiments}

\subsection{Datasets and Evaluation Settings}
We use 3 popular datasets (SCUT-FBP~\cite{xie2015scut}, HotOrNot~\cite{gray2010predicting} and SCUT-FBP5500~\cite{liang2018scut}) to evaluate the effectiveness of our proposed methods.
SCUT-FBP~\cite{xie2015scut} contains 500 female images with beauty scores ranged from 1 to 5. HotOrNot dataset~\cite{gray2010predicting} contains 2056 facial images collected from Internet with variant postures, cluttered background, overexposure and low resolution problems. SCUT-FBP5500~\cite{liang2018scut} contains 5500 facial images with beauty scores in $[1, 5]$. Each image is annotated by 60 volunteers, and the average is used as the groundtruth to remove personal preference bias. All experiments are conducted by 5-fold cross validation, and the average is reported for comparison with other models.

As noted in related works~\cite{xie2015scut,gray2010predicting,liang2018scut}, we adopt mean absolute error (MAE), root mean squared error (RMSE) and pearson correlation (PC) to evaluate the performance of different models on SCUT-FBP5500~\cite{liang2018scut}, and PC is used as metric on SCUT-FBP~\cite{xie2015scut} and HotOrNot~\cite{gray2010predicting} datasets. A computational model with lower MAE, lower RMSE and higher PC denotes better performance.

\subsection{Implementation Details}
We conduct experiments with PyTorch~\cite{paszke2019pytorch} on two NVIDIA K80 GPUs with cuDNN acceleration. The learning rate starts from 0.01 and is divided by 10 per 50 epochs. Weight decay and batch size are set as 0.001 and 64, respectively. The model is trained via SGD with 0.9 momentum for 200 epochs. The images are resized and randomly cropped to $224\times 224$ patches, color jittering and random rotation~\cite{paszke2019pytorch} are applied for data augmentation. The network is initialized with ImageNet pretrained weights to avoid overfitting due to the limited capacity of current public research dataset. We do not adopt additional facial dataset for pretraining.

\subsection{Performance Evaluation}

\begin{figure}[thbp]
	\includegraphics[width=0.38\textheight]{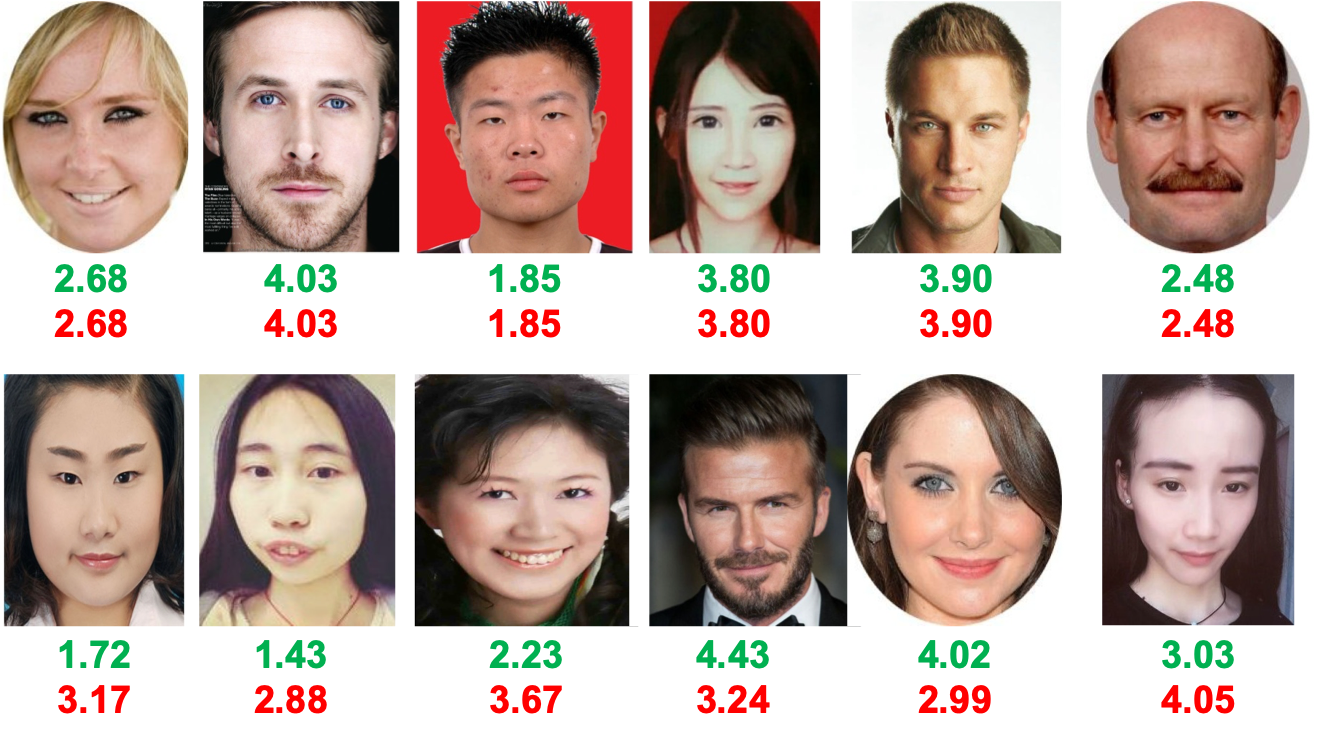}
	\centering
	\caption{Precisely prediction and inaccurately output samples, \textcolor{green}{green} and \textcolor{red}{red} scores represent groundtruth score and predicted score, respectively.}
	\label{fig:samples}
\end{figure}

The performance comparison with other methods on SCUT-FBP~\cite{xie2015scut}, HotOrNot~\cite{gray2010predicting} and SCUT-FBP5500~\cite{liang2018scut} datasets are shown in Table~\ref{tab:scutfbp_results}, Tabel~\ref{tab:hotornot_results} and Table~\ref{tab:scutfbp5500_results}, respectively. The proposed ComboLoss achieves state-of-the-art performance and outperforms prior arts on SCUT-FBP~\cite{xie2015scut}, HotOrNot~\cite{gray2010predicting} and SCUT-FBP5500~\cite{liang2018scut} datasets by 1.13\%, 2.1\% and 0.57\% on PC, respectively. We list 6 precisely predicted samples and 6 inaccurately output instances on SCUT-FBP5500~\cite{liang2018scut} in Fig~\ref{fig:samples}.

\begin{table}[htbp]
	\centering
	\caption{Performance comparison on SCUT-FBP.}
	\label{tab:scutfbp_results}
	\begin{tabular}{|c|c|}
		\hline
		\textbf{Model} & \textbf{PC}\\\hline
		Combined Features+Gaussian Regression~\cite{xie2015scut} & 0.6482\\\hline
		CNN-based~\cite{xie2015scut} & 0.8187\\\hline
		Liu et al.~\cite{liu2017landmark} & 0.6938\\\hline
		Xu et al.~\cite{xu2018transferring} & 0.8570\\\hline
		KFME~\cite{elorza2017face} & 0.7988\\\hline
		RegionScatNet~\cite{liu2017facial} & 0.83\\\hline
		PI-CNN~\cite{Xu2017Facial} & 0.87\\\hline
		CRNet~\cite{xu2018crnet} & 0.8723\\\hline
		HMTNet + Ridge Regression~\cite{xu2019hierarchical} & 0.8977 \\\hline
		\textbf{ComboLoss (Ours)} & \textbf{0.9090}\\\hline
	\end{tabular}
\end{table}

\begin{table}[htbp]
	\centering
	\caption{Performance comparison on HotOrNot Dataset.}
	\label{tab:hotornot_results}
	\begin{tabular}{|c|c|}
		\hline
		\textbf{Model} & \textbf{PC}\\\hline
		Eigenface~\cite{gray2010predicting} & 0.180\\\hline
		Two Layer Model~\cite{gray2010predicting} & 0.438\\\hline
		Multiscale Model~\cite{gray2010predicting} & 0.458 \\\hline
		S. Wang et al.~\cite{wang2014attractive} & 0.437 \\\hline

		Xu et al.~\cite{xu2018transferring} & 0.468\\\hline
		CRNet~\cite{xu2018crnet} &0.482 \\\hline
		\textbf{ComboLoss (Ours)} & \textbf{0.503}\\\hline
	\end{tabular}
\end{table}

\begin{table}[htbp]
	\centering
	\caption{Performance comparison on SCUT-FBP5500.}
	\label{tab:scutfbp5500_results}
	\begin{tabular}{|c|c|c|c|}
		\hline
		\textbf{Model} & \textbf{MAE} & \textbf{RMSE} & \textbf{PC}\\\hline
		ResNeXt-50~\cite{xie2017aggregated} & 0.2291 & 0.3017 & 0.8997\\\hline
		ResNet-18~\cite{he2016deep} & 0.2419 & 0.3166 & 0.8900\\\hline
		AlexNet~\cite{krizhevsky2012imagenet} & 0.2651 & 0.3481 & 0.8634\\\hline
		HMTNet~\cite{xu2019hierarchical} & 0.2380	& 0.3141 & 0.8912\\\hline
		AaNet~\cite{lin2019attribute} & 0.2236 & 0.2954 & 0.9055\\\hline
		$R^2$ ResNeXt~\cite{lin2018r} & 0.2416 & 0.3046 & 0.8957\\\hline
		$R^3$CNN~\cite{lin2019regression} & 0.2120 & 0.2800 & 0.9142\\\hline
		\textbf{ComboLoss (Ours)} & \textbf{0.2050} & \textbf{0.2704} & \textbf{0.9199}\\\hline
	\end{tabular}
\end{table}

\subsection{Ablation Study}
To validate the effectiveness of each component of our proposed methods, we perform extra ablation experiments on SCUT-FBP5500 dataset~\cite{liang2018scut}. Unlike the experimental settings used in the previous performance comparison section, we take another data splitting strategy as defined in SCUT-FBP5500 provision~\cite{liang2018scut}. Namely, the 60\% of the images are used as training set and the remaining images are test set. We adopt 60\%/40\% splitting setting~\cite{liang2018scut} in ablation experiments to reduce huge computation of 5-fold cross validation. 

\subsubsection{Effects on ComboLoss}
We conduct experiments under the supervision of different loss functions, namely, MSE loss~\cite{xie2015scut,liang2018scut}, $L_1$ loss, smooth $L_1$ loss and Smooth huber loss~\cite{xu2019hierarchical} (see Table~\ref{tab:loss_comp}). ComboLoss achieves best performance, and surpasses MSE loss on MAE, RMSE and PC by 0.69\%, 1.34\% and 1.09\%, respectively.

\begin{table}[!bhtp]
	\centering
	\caption{Evaluation on different loss functions.}
	\label{tab:loss_comp}
	\begin{tabular}{|c|c|c|c|}
		\hline
		\textbf{Loss Function} & \textbf{MAE} & \textbf{RMSE} & \textbf{PC}\\\hline
		$L_1$ Loss & 0.2191	& 0.2918 & 0.9030\\\hline
		MSE Loss &  0.2195 &	0.2947 & 0.9008\\\hline
		Smooth $L_1$ Loss & 0.2194 & 0.2869 & 0.9064\\\hline
		Smooth Huber Loss~\cite{xu2019hierarchical} & 0.2196 & 0.2903 & 0.9052\\\hline
		\textbf{ComboLoss (Ours)} & \textbf{0.2126} & \textbf{0.2813} & \textbf{0.9117}\\\hline
	\end{tabular}
\end{table}

\subsubsection{Effects of Different Network Backbone}
Backbone network plays a vital part in feature representation learning and performance~\cite{krizhevsky2012imagenet,he2016deep,xie2017aggregated,hu2018squeeze,tan2019efficientnet}. We replace the SEResNeXt50~\cite{hu2018squeeze} with a simple ResNet18~\cite{he2016deep} to validate the effectiveness of backbone architecture and supervision of ComboLoss. We can see clearly from Table~\ref{tab:arch} that stronger backbone architecture leads to better performance (0.9041 VS 0.9117 and 0.8946 VS 0.9008). However, the superior performance our proposed methods achieved does not solely come from stronger backbone. Supervised by the vanilla MSE loss, the stronger SEResNeXt50~\cite{hu2018squeeze} only achieves a PC of 0.9008, while a simple ResNet18~\cite{he2016deep} supervised by ComboLoss achieves a PC of 0.9041. The SEResNeXt50 trained by ComboLoss result in 1.09\% performance gain than its counterpart SEResNeXt50 trained by MSE Loss, which shows the effectiveness of the proposed methods as well.

\begin{table}[bthp]
	\centering
	\caption{Evaluation on different backbone architecture.}
	\label{tab:arch}
	\begin{tabular}{|c|c|c|c|}
		\hline
		\textbf{Model} & \textbf{MAE} & \textbf{RMSE} & \textbf{PC}\\\hline
		ResNet18 & 0.2313 & 0.3054 & 0.8946 \\\hline
		ResNet18 + ComboLoss & 0.2202 & 0.2907	& 0.9041 \\\hline
		SEResNeXt50 &  0.2195 &	0.2947 & 0.9008  \\\hline
		SEResNeXt50 + ComboLoss  & \textbf{0.2126} & \textbf{0.2813} & \textbf{0.9117}\\\hline
	\end{tabular}
\end{table}

\section{Conclusion and Future Work}
In this paper. we first perform systematic analysis on diverse loss functions for facial attractiveness score regression. Then a simple yet effective approach named ComboLoss, is presented to enhance model training. The proposed method achieves state-of-the-art performance on SCUT-FBP~\cite{xie2015scut}, HotOrNot~\cite{gray2010predicting} and SCUT-FBP5500~\cite{liang2018scut} datasets. We will apply ComboLoss to other regression problems (such as age estimation), and explore more useful discretization approaches in our future work.
\newpage

\bibliographystyle{IEEEbib}
\bibliography{refs}

\end{document}